\documentclass[21,reqno]{amsart}

\usepackage[english, activeacute]{babel}
\usepackage{amsmath,amsthm,amssymb}
\usepackage{epsfig}
\usepackage{latexsym}
\usepackage{amsfonts}
\usepackage[active]{srcltx}
\usepackage{graphics}
\usepackage{eepic}
\usepackage{epic}
\usepackage{pstricks,pst-grad}

\title{Dynamical mobility edge for various random Landau Hamiltonians}
\author{F. \textsc{Germinet} \and C. \textsc{Rojas-Molina}}
\address{Universit\'e de Cergy-Pontoise, IUF, UMR CNRS 8088, F-95000 Cergy-Pontoise,
France.} 
\email{francois.germinet@u-cergy.fr}
\address{Universit\'e de Cergy Pontoise, UMR CNRS 8088, F-95000 Cergy-Pontoise,
France.}
\email{constanza.rojas-molina@u-cergy.fr}

\newtheorem{thm}{Theorem}[section]

\newtheorem{lmm}[thm]{Lemma}

\theoremstyle{definition}

\theoremstyle{remark}

\newcommand{\be}{\begin{equation}}
\newcommand{\ee}{\end{equation}}
\newcommand{\ba}{\begin{array}}
\newcommand{\ea}{\end{array}}
\newcommand{\bal}{\begin{align}}
\newcommand{\eal}{\end{align}}
\newcommand{\bea}{\begin{eqnarray}}
\newcommand{\eea}{\end{eqnarray}}
\newcommand{\bee}{\begin{eqnarray*}}
\newcommand{\eee}{\end{eqnarray*}}

\renewcommand{\P}{\mathbb P}
\newcommand{\Z}{\mathbb Z}
\newcommand{\Rd}{\mathbb R^d}
\newcommand{\R}{\mathbb R}

\newcommand{\abs}[1]{\left| #1 \right|}

\renewcommand{\L}{\Lambda}

\newcommand{\rd}{{\mathbb R}^{2}}
\newcommand{\eq}[1]{\eqref{#1}}
\newcommand\beq{\begin{equation}}
\newcommand\eeq{\end{equation}}
\def\I{\mathcal{I}}
\def\eps{\varepsilon}

\newcommand{\G}{\mathbb{G}}

\DeclareMathOperator{\supp}{supp}

\newcommand{\la}{\langle}
\newcommand{\ra}{\rangle}
\newcommand{\scal}[1]{\la #1 \ra}

\begin{document}

\maketitle 

\begin{abstract}      

We review recent results obtained within the framework of the integer quantum Hall effect in the spirit of the work of Germinet, Klein, Schenker in \cite{GKS}. Landau Hamiltonians perturbed by random electric or magnetic perturbations are shown to exhibit a dynamical mobility edge, that is a transition between a regime of dynamical localization and a regime of non trivial transport at a minimal rate. The focus is put on three situations of interest: 1) unbounded ergodic electric potentials, for which Landau gaps are filled; 2) non ergodic electric potentials; 3) random magnetic potentials.
\end{abstract}

\section{Introduction}

Random Schr\"odinger operators appear in a natural way within the theory of integer quantum Hall effect, for impurities are responsible for the occurrence of the famous ``plateaux" between two jumps of the Hall conductance, as pointed out by Bellissard \cite{Be}. This phenomenon discovered by von Klizting et alii \cite{vKli} has the particularity of being very robust. It can be described as follows: energies between two successive Landau levels are trapped so that the direct conductance vanishes and the Hall conductance is constant, while near the Landau levels wave-packets are expected to be delocalized. In particular, that dynamical localization fails near landau levels has been shown by \cite{BES} in the discrete setting.

For Anderson type perturbations of the Landau Hamiltonian in $\R^2$, the existence of a dynamical mobility edge has been proved by Germinet, Klein, Schenker in \cite{GKS}. A strong form of dynamical delocalization is obtained, and wave-packets are shown to travel at a given minimal speed (through a lower bound on high enough order moments) and for some energies that are asymptotically close to the Landau levels as the magnetic strength is large or the disorder is small.

The core of the proof relies on two main ingredients: the characterization of the region of dynamical localization of \cite{GK3} where slow transport is shown to be absent, and the constancy of the Hall conductance in the region of dynamical localization. In particular we point out that the famous ``Anderson localization" property, namely exponential decay of the eigenfunctions \cite{An}, is not strong enough to guarantee the constancy of the Hall conductance, and one requires more detailed informations on the eigenfunctions behaviour.

In \cite{GKS2}, the result is extended to more general ergodic models for which the Hall conductance is shown to be integer valued. In \cite{GKM} unbounded electric potentials are considered, in which case the Landau gaps do not survive, as soon as the disorder is turned on. In \cite{RM} non ergodic random electric potentials are studied. In \cite{DGR}, the result is extended to random magnetic potentials. 

The fact that the same dynamical transition phenomenon holds in these quite different situations can be interpreted as another indication of the robustness of the integer quantum Hall effect.

Let us now describe the models and the typical type of results one obtains.
    Let ${\bf A} = (A_1,A_2) \in L_{\rm loc}^2(\rd,
\rd)$ be a magnetic potential. Define the operator $H({\bf A})$ as
the self-adjoint operator generated in $L^2(\rd)$ by the closure
of the quadratic form
$$
\int_{\rd} |i\nabla u + {\bf A}u|^2 dx, \quad u \in C_0^{\infty}(\rd).
$$
The magnetic field generated by ${\bf A}$ is
$$
B : = \frac{\partial A_2}{\partial x_1} - \frac{\partial
A_1}{\partial x_2}.
$$
When the magnetic field is a constant $B>0$,  the operator is just the well-known Landau Hamiltonian, which we will denote by $H({\bf A}_0)$, where $A_0$ generates the constant magnetic field $B$.  It is well-known
that the spectrum  $\sigma(H({\bf A}_0))$  of the  Landau
Hamiltonian $H({\bf A}_0)$ consists of a sequence of infinitely
degenerate eigenvalues, the so called
Landau levels:  { 
\begin{equation} \label{landaulevels}
B_n=(2n-1)B ,\quad n=1,2,\dotsc .
\end{equation}}
For further reference, we also set  {
\beq
\mathcal{B}_1=]-\infty,2B[,\quad \mathrm{and} \quad \mathcal{B}_n=]B_n-B,B_n+B[, \quad n=2,3,\dotsc .
\eeq}
We shall study random perturbations of $H({\bf A}_0)$, either magnetic or electric. More precisely we consider  operators of the form
\begin{equation} \label{defHelec} 
H_{B,\lambda,\omega} =H({\bf A}_0) +
\lambda  V_{\omega} \quad \mathrm{on} \quad
{L}^2(\mathbb{R}^2, {\mathrm{d}}x), 
\end{equation}
The parameter $\lambda \ge 0$ is
the disorder parameter,   and the electric perturbation
$V_\omega$ is a random potential of the form
\begin{equation}\label{defV}
V_\omega(x) =\sum_{\gamma\in\Gamma} \omega_\gamma u_\gamma,
\end{equation}
where $\Gamma$ is a countable subset of $\R^2$ (typically a lattice or a Delone set) and $u_\gamma=u(x-\gamma)$; the single site potential $u$ is a  nonnegative bounded 
measurable function
on $\R^{d}$ with compact support, uniformly 
bounded away from zero in
a neighborhood of the origin;   the $\omega_j$'s are  independent, identically distributed  random variables,  whose common  probability distribution $\mu$ has  a bounded density $\rho$.
We fix constants for $u$ by 
\begin{equation} 
C_{-}\chi_{\Lambda_{\delta_{-}}(0)}\le u \le C_+ \chi_{\Lambda_{\delta_{+}}(0)}\quad \text{with $C_{\pm}, \delta_{\pm}\in ]0,\infty[ $},
\end{equation}
and normalize $u$ so that we have  $\|\sum_{\gamma\in \Gamma} u_j\|_\infty\le 1$.

Next, we consider
\begin{equation} \label{defHmagn} 
H_{B,\lambda,\omega} =H({\bf A}_0 + \lambda{\bf A}_\omega) , \quad \mathrm{on} \quad
{L}^2(\mathbb{R}^2, {\mathrm{d}}x),
\end{equation}
where  the magnetic perturbation ${\bf A}_\omega$ is given by a random magnetic potential of the form
    \beq\label{tt1}
{\bf A}_{\omega}(x)=  \sum_{\gamma\in \Z^2} \omega_\gamma {\bf
u}_\gamma(x),
    \eeq
with ${\bf u}_\gamma(x)=(u_1(x -\gamma),u_2(x -\gamma))$, $\gamma
\in {\mathbb Z}^2$, $x \in \R^2$, $u_1,u_2$ being two given
$C^1(\R^2,\R)$ compactly supported functions, normalized so that
$\|\sum_{\gamma\in\Z^2} {\bf u}_\gamma\|_\infty\le 1$; as above the random
variables $(\omega_\gamma)_{\gamma\in\Z^2}$ are independent and
identically distributed, supported on $[-1,1]$, with common
density $\rho$.

To investigate the localization and delocalization properties of such operators, we study the time spreading of wave-packets initially localized in space and energy at time $t=0$.

Following \cite{GK3,GKjsp,GKS,GKS2}, we set $\Xi_{B,\lambda}^{\mathrm{DL}}$ to be the region of complete localization (gaps included), that is, the set of energies where  the multiscale analysis applies or the fractional moment method of \cite{AENSS}. Its complement  is the set of dynamical delocalization $\Xi_{B,\lambda}^{\mathrm{DD}}$. An energy $E\in\Xi_{B,\lambda}^{\mathrm{DD}}$ such that for any $\eps>0$, $[E-\eps,E+\eps]\cap\Xi_{B,\lambda}^{\mathrm{DL}}\not=\emptyset$, is called a dynamical mobility edge.

As shown in  \cite{GK3}, the region of complete localization  $\Xi_{B,\lambda}^{\mathrm{DL}}$ can be characterized as the region of dynamical localization. To measure `dynamical localization' we introduce
\begin{equation}\label{moment}
M_{B,\lambda,\omega}(p,\mathcal{X},u,t)  = 
\left\|  {\langle} x-u {\rangle}^{\frac p 2}
{\mathrm{e}^{-i tH_{B,\lambda,\omega} }}
\mathcal{X}(H_{B,\lambda,\omega}) \chi_u 
\right\|_2^2 ,
\end{equation}
the
random moment of order
$p\ge 0$ at time $t$ for the time evolution  in the Hilbert-Schmidt norm,  
initially spatially localized in the square of side one around some origin $u\in\R^2$
(with characteristic function $\chi_u$), and ``localized" 
in energy by the function $\mathcal{X}\in C^\infty_{c,+} (\mathbb{R})$.
Its time averaged expectation is given by
\begin{equation}   \label{tam}
\mathcal{M}_{B,\lambda}( p ,\mathcal{X}, T )   =  \sup_{u\in\R^2}
\frac1{T} \int_0^{\infty}
\mathbb{E}\left\{ M_{B,\lambda,\omega}(p,\mathcal{X},u,t)\right\}
{\mathrm{e}^{-\frac{t}{T}}} \,{\rm d}t .
\end{equation}
Note that in the ergodic situation, $\mathbb{E}\left\{ M_{B,\lambda,\omega}(p,\mathcal{X},u,t)\right\}=\mathbb{E}\left\{ M_{B,\lambda,\omega}(p,\mathcal{X},0,t)\right\}$ for any $u\in\R^2$, so that it is enough to consider $u=0$. But when translation invariance is lost, definitions like \eqref{moment}-\eqref{tam} are required.

It is proven in  \cite{GK3}, for ergodic models, that $ \Xi_{B,\lambda}^{\mathrm{DL}}$ coincides with the set of energies $E$ for which there exists 
 $\mathcal{X}\in C^\infty_{c,+} (\mathbb{R})$ with $\mathcal{X}\equiv 1$ on some open interval containing $E$, $\alpha \ge 0$, and $p > 4\alpha + 22$, such that
 \beq \label{slowdec}
 \liminf_{T \to \infty} \frac 1 {T^\alpha}\mathcal{M}_{B,\lambda}( p ,\mathcal{X}, T ) < \infty,
 \eeq
in which case it is also shown in \cite{GK3} that  \eq{slowdec} holds for any $p\ge 0$ with $\alpha=0$.

The typical results that are proved for Hamiltonians of the form \eqref{defHelec} and \eqref{defHmagn} read as follows.

\begin{thm}\label{typicalthm}
Given $N$, for suitable parameters $B$, $\lambda$, operators $H_{B,\lambda,\omega}$ exhibit dynamical localization and delocalization in each Landau band  $\mathcal{B}_n$, $n=1,\cdots,N$, that is, for any  $n=1,\cdots,N$,
\begin{equation} 
\Xi_{B,\lambda}^{\mathrm{DL}} \cap \mathcal B_n\neq
 \emptyset \quad \mbox{ and } \quad
\Xi_{B,\lambda}^{\mathrm{DD}} \cap \mathcal B_n\neq \emptyset. \label{set}
\end{equation}
In particular, there exists dynamical mobility edges $E_{j,n}(B,\lambda)\in
 \mathcal B_n$, $j=1,2$ (it is possible that $E_{1,n}(B,\lambda)=E_{2,n}(B,\lambda)$).
\end{thm}

Since $E\in\Xi_{B,\lambda}^{\mathrm{DD}}$ means dynamical delocalization in the sense  that \eq{slowdec} does not hold for any
 $\mathcal{X}\in C^\infty_{c,+} (\mathbb{R})$ with $\mathcal{X}\equiv 1$ on some open interval containing $E$, $\alpha \ge 0$, and $p > 4\alpha + 22$, Theorem~\ref{typicalthm} has the following consequence in terms of transport properties of the random Landau Hamiltonian $H_{B,\lambda,\omega}$

\begin{thm} Given $N$, for suitable parameters $B$, $\lambda$, the random Landau Hamiltonian  
$H_{B,\lambda,\omega}$ 
exhibits   dynamical delocalization
in each Landau band $ \mathcal{B}_n$, \linebreak  $n= 1,\cdots,N$: 
there exists at least one energy 
$E_n(B,\lambda)\in \mathcal{B}_n $,
such that for every  
$\mathcal{X}\in \mathcal{C}^\infty_{c,+} (\mathbb{R})$  with
$\mathcal{X} \equiv 1$  on some open interval  $J\ni E_n(B,\lambda) $
and  $p>0$,  we have 
\begin{equation}\label{momentgrowth}
\mathcal{M}_{B,\lambda}(p,\mathcal{X}, T)    \ge \
C_{p,\mathcal{X}} \, T^{\frac p4 - 6} \ ,
\end{equation}
for all  $T \ge 0$  with  $  C_{p,\mathcal{X}} > 0 $.
\end{thm}

As mentioned at the beginning, the core of the proof goes back to \cite{GKS} for ergodic bounded potentials  of the form \eqref{defV} and relies on two main ingredients: the characterization of the region of dynamical localization of \cite{GK3} and the constancy of the Hall conductance in the region of dynamical localization.

In \cite{GKM}, where unbounded electric potentials are considered, the analysis of the properties of the Hall conductance relies on the work \cite{GKS2}, for one needs to know a priori that the Hall conductance is integer valued in the region of dynamical localization.

In \cite{RM} the random electric potential is designed using a Delone underlying set, so that ergodicity is lost. In particular the analysis of \cite{GKS2} fails and one rather uses the more direct approach of \cite{GKS}. The analog of the  characterization of the region of dynamical localization of \cite{GK3} is established in \cite{RM}.

In \cite{DGR}, the result is extended to random magnetic potentials, a challenging class of models, for which localization is not yet widely established (see however \cite{GHK}).

We point out that the location of the spectrum is part of the study of the model in order to make sure that statements in \eqref{set} are not empty. 

In the remaining part of this note, we shall review the results of these three last works: \cite{GKM} in Section \ref{sectubd}, \cite{RM} in Section \ref{sectnonerg} and \cite{DGR} in Section \ref{sectmagn}.


\section{Unbounded ergodic random electric potential}
\label{sectubd}

 In this section, we review the result of \cite{GKM} where the density $\rho$  is taken  $\supp \rho= \R$. The function $\rho$ is assumed to satisfy a fast decay property:
\beq \label{rhodecay}
\rho(\omega) \le \rho_0 \exp(-|\omega|^\alpha),
\eeq
for some $\rho_0\in]0,+\infty[$ and $\alpha>0$.  Under these hypotheses,  $H_{B,\lambda,\omega}$ is  essentially self-adjoint on $\mathcal{C}_c^\infty(\R^d)$ with probability one,
with the bound  ($\scal{x}:=\sqrt{1 +\abs{x}^2})$
\begin{equation}
H_{B,\lambda,\omega}\ge - c_\omega (\log \scal{x})^\beta, \mbox{ for all } x\in\R^d,
\end{equation} 
for any given $\beta>\alpha^{-1}$, { with} $c_\omega$ depending also on $\alpha,\beta,d$.

The particularity of this model is that, as soon as $\lambda>0$, the spectrum fills the Landau gaps and we have 
\beq
\sigma(H_{B,\lambda,\omega})=\R, \quad \P-\mbox{a.s.}
\eeq
The fact that the Landau gaps are immediately filled up as soon as the disorder is turned on implies that  the approach used in \cite{GKS} is  non applicable, and the authors resorts to the full theory developed in \cite{GKS2}, which is the analog of \cite {BES,AG} but in a  continuum setting.

\begin{thm}[\cite{GKM}]\label{thmlimit2}   Let  $H_{B,\lambda,\omega}$ be a  random Landau Hamiltonian   as above.
For
each $n=1,2,\dots$, if $\lambda$ is small enough (depending on $n$) there exist dynamical mobility edges $E_{j,n}(B,\lambda)\in\mathcal{B}_n$, 
$j=1,2$, such that
\begin{align} 
\max_{j=1,2} \left \lvert E_{j,n}(B,\lambda)  - B_n  \right \rvert  
\le  K_n(B)\lambda
\abs{\log \lambda}^{\frac 1 \alpha} \to  0 \quad 
\text{as $\lambda \to 0$},
\label{lambda21}
\end{align}
with a finite  constant  $ K_n(B) $. 
(It is possible
that  ${E}_{1,n}(B,\lambda)=
{E}_{2,n}(B,\lambda)$, i.e., dynamical delocalization occurs at 
a single energy.)
\end{thm}


\section{Non ergodic random electric potential}
\label{sectnonerg}

In this section we present results obtained in \cite{RM} for Landau Hamiltonians with non ergodic potentials. Consider the random operator as in \eqref{defHelec}, namely
\begin{equation} 
H_{B,\lambda,\omega}=(-i\nabla-{\bf A_0})^2+\lambda V_\omega,
\end{equation}
where this time $V_\omega$ is a Delone-Anderson potential, defined by
\begin{equation}\label{ranpot}
V_\omega(x)=\displaystyle\sum_{\gamma \in D}\omega_\gamma
u(x-\gamma),
\end{equation}
where $D$ is a Delone set, i.e. a uniformly discrete and relatively dense set in ${\mathbb R}^2$, not necessarily periodic and $(\omega_\gamma)$ are i.i.d random variables with common probability density $\rho$ and $\mbox{supp }\rho=[-m_0,M_0]$. Locations of obstacles being irregular, ergodicity is lost. 

We  assume furthermore that the Landau bands $\mathcal B_{n}(B,\lambda)$ containing the spectrum of $H_{B,\lambda,\omega}$ are disjoint, that is
\[\lambda(m_0+M_0)<2B\]

\begin{thm}[\cite{RM}]
For each  $n=1,2,\dots$, if $B$ is large enough (depending on $n$) there exist dynamical mobility edges 
${E}_{j,n}(B,\lambda)$, 
$j=1,2$, with
\begin{gather}  \max_{j=1,2} \left \lvert {E}_{j,n}(B,\lambda)  - B_n  \right \rvert  
\le  K_n(\lambda)\frac{\log B}  B  \to  0 \quad \text{as $B \to \infty$},
\label{B2infty}
\end{gather}
where $K_n(\lambda) $ denotes a finite constant.  (It is possible
 that  ${E}_{1,n}(B,\lambda)=
{E}_{2,n}(B,\lambda)$, i.e., dynamical delocalization occurs at 
a single energy.)
\end{thm}

To get such a result, one has to extend the bootstrap multiscale analysis of \cite{GK1} as well as the characterization of the Anderson metal-insulator transition of \cite{GK3}. Next we take advantage of the approach of \cite{GKS} to prove the transition, for it does not require ergodicity (note that results from \cite{GKS2} are not applicable here).

We say $H_{\omega}$ exhibits strong Hilbert-Schmidt (HS) dynamical
localization in the open interval $I$ if for all $\mathcal X\in\mathcal
C^\infty_{c,+}(I)$ we have
\[ \displaystyle
\sup_{u\in\mathbb Z^2}\mathbb E\left( \sup_{t\in \mathbb R}  M_{u,\omega}(p,\mathcal X, t)\right)<\infty \hspace{0.5cm}
\mbox{for all $p\geq 0$}  \]
$H_{\omega}$ exhibits strong HS-dynamical localization at an energy $E$ if there exists an  open interval $I$ with $E\in
I$, such that there is strong HS-dynamical localization in the open interval.
Next we define the region of complete localization for $H_\omega$ as
\begin{equation}
\Sigma_{CL}= \{E\in\mathbb R: \mbox{$H_{\omega}$ exhibits strong HS-dynamical
localization at $E$} \}
\end{equation}
Note that if $E\in \mathbb R\setminus\sigma_\omega$ for a.e. $\omega$, then $E\in\Sigma_{CL}$.

To state the result we need the following definitions.
Given $\theta>0$, $E\in\mathbb R$, $x\in\mathbb Z^d$ and $L\in 6\mathbb N$, we
say that the box $\L_L(x)$ is $(\theta,E)$-\emph{suitable}  for $H_\omega$
if $E\notin\sigma_{\omega,x,L}$ and
\[  \Vert \Gamma_{x,L}
R_{\omega,x,L}(E)\chi_{x,L/3}\Vert_{x,L}\leq \frac{1}{L^\theta} , \]
where $\Gamma_{x,L}=\chi_{\bar\L_{L-1}(x)\setminus\L_{L-3}(x)}$. 

Next, we say that $H_\omega$ satisfies a uniform Wegner estimate with H\"older
exponent s in an open interval $\mathcal J$ if for every $E\in\mathcal J$
there exists a constant $Q_E$, bounded on compact subintervals of $\mathcal J$
and $0<s\leq 1$ such that 
\begin{equation}\label{WE} \sup_{x\in\Rd}\mathbb P\{ \mbox{dist}(\sigma_{\omega,x,L},
E)\leq \eta \}\leq Q_E\eta^s L^d
\end{equation}
for all $\eta>0$ and $L\in2\mathbb N$.  It satisfies a uniform Wegner estimate
at an energy $E$ if it satisfies a uniform Wegner estimate in an open interval
$\mathcal J$ such that $E\in\mathcal J$.

The following is a reformulation of the Bootstrap Multiscale Analysis (MSA) of Germinet and Klein {\cite{GK1}} in the non ergodic setting.

\begin{thm}[\cite{RM}]\label{Bootstrap}
Assume $H_\omega$ satisfies a uniform Wegner estimate with H\"older exponent \emph{s} .  Given $\theta>d/s$, for each $E\in\mathcal J$ there
exists a finite scale $\mathcal L_\theta(E)=\mathcal L(\theta,E,Q_E,d,s)$,
bounded in compact subintervals of $\mathcal J$, such that if for $\mathcal L>\mathcal L_\theta(E)$ the following holds
\begin{equation}
\label{ILSE} \inf_{x\in\mathbb Z^d} \mathbb P\{\L_{\mathcal L}(x) \mbox{
is ($\theta,E$)-suitable}\}> 1-\frac{1}{841^d},
\end{equation}
then there exists $\delta_0>0$ and $C_\zeta>0$ such that
\begin{equation}\sup_{u\in\mathbb Z^d} \mathbb E \left(  \displaystyle\sup_{\Vert
f\Vert\leq 1} \Vert \chi_{x+u} f(H_\omega) E_\omega(I(\delta_0))\chi_u\Vert_2^2
\right)\leq C_\zeta e^{-|x|^\zeta}, 
\end{equation}
for $0<\zeta <1$, where $I(\delta_0)=[E-\delta_0,E+\delta_0]$.  Moreover, $E\in
\Sigma_{CL}$.
\end{thm}

We define the multiscale analysis region for $H_{\omega}$ as the set of
energies where we can perform the bootstrap MSA, i.e.

\begin{align}\Sigma_{MSA}= & \{E\in\mathbb R: \mbox{$H_{\omega}$ satisfies a
uniform Wegner estimate at $E$ and}\nonumber\\ &   \mbox{ (\ref{ILSE}) holds for some  $\mathcal L>\mathcal
L_\theta(E)$} \}
\end{align}
By Theorem \ref{Bootstrap}, we have  $\Sigma_{MSA}\subset\Sigma_{CL}$.

The following result is an extension of Theorem 2.11 \cite{GK3},  for the non
ergodic setting in annealed regimes

\begin{thm}[\cite{RM}]\label{Emom} Let $H_{\omega}$ be as above.  Let
$\mathcal X \in C^\infty_{c,+}(\mathbb R)$ with $\mathcal X \equiv
1$ on some open interval $J\subset \mathcal J$, $\alpha\geq0$ and
$p>p(\alpha,s):= 12 \frac{d}{s}+2\alpha\frac{d}{s}$.  If
\begin{equation} \label{Emomentum} \displaystyle\liminf_{T\rightarrow\infty}
\sup_{u\in\mathbb Z^d} \frac{1}{T^{\alpha}}\mathbb E \left(\mathcal M_{u,\omega}(p,\mathcal X,T)\right)< \infty ,\end{equation}
then $J\subset\Sigma_{MSA}$. In particular, it follows that
(\ref{Emomentum}) holds for any $p\geq0$.
\end{thm}

Moreover, \cite{RM} strengthens this result by improving it in a quenched regime.

\begin{thm}\label{mom} Let $H_{\omega}$, $\mathcal X $ and $\alpha$ be as above and
$p>p(\alpha,s):= 15 \frac{d}{s}+2\alpha\frac{d}{s}$.  If
\begin{equation}\label{momentum} \displaystyle\liminf_{T\rightarrow\infty}
\sup_{u\in\mathbb Z^d}T^{\frac s 2}\mathbb P (\mathcal M_{u,\omega}(p,\mathcal X,T)>
T^\alpha)=0,\end{equation}
then $J\subset\Sigma_{MSA}$. In particular, it follows that
(\ref{momentum}) holds for any $p\geq0$.
\end{thm}

The last theorem relies on the following lemma that makes the link between a slow transport property of the dynamics and  the uniform initial length scale estimate we need to start the adapted MSA.

\begin{lmm}\label{deterministiclemma}
Let $\theta>d/s$ and $\gamma>d/s$.  There exists $\mathcal L =\mathcal L(I,p,\theta,\gamma,d,\alpha,s,p_0, Q_I)$ such that for any $u\in \tilde\L_{L/3}(y)$ with $L(I,\epsilon)\geq \mathcal L$ and $E\in I$ fixed, if 
\begin{equation} p>p(\theta,\gamma,d,\alpha,s):= \alpha\frac{(\theta s+d)}{s}+9\theta+3\gamma+2d+\frac d s\end{equation}
then, for $T=\epsilon^{-1}$,
\begin{equation}\left\{ \omega:\mbox{ }\Vert
\Gamma_{y,L} R_\omega(E+i\epsilon)\mathcal
X(H_\omega)\chi_{u}\Vert>\frac{1}{2L(I,\epsilon)^{\theta+\gamma+d}}\right\}\subset \left\{\omega:\mbox{ }\mathcal M_{u,\omega}(p,\mathcal X,T)>T^\alpha\right\} .
\end{equation}
\end{lmm}


\section{Ergodic random magnetic potential}
\label{sectmagn}

In this section, we review the result of \cite{DGR}. The Hamiltonian is of the form \eqref{defHmagn}, that is the randomness occurs through  the magnetic potential only. More precisely, we denote by
$$
H_{B,\lambda,\omega,\eta} : = H({\bf A}_0 + \lambda {\bf
A}_{\omega,\eta})
$$
the corresponding magnetic random operator with a common density of the random variables
$$
\rho_\eta(s) ds = C_{\eta} \eta^{-1} \exp(-|s| \eta^{-1})
\chi_{[-1,1]}(s), \quad \eta>0,
$$
 and $C_\eta$ such that $\int
\rho_\eta ds =1$ (note that $\frac12\le C_\eta \le 1$ for
$\eta\in]0,1]$). The support of $\rho_\eta$ is $[-1,1]$ for all
$\eta>0$, but as $\eta$ goes to zero, the disorder becomes weaker
in the sense that for most $\gamma$ the coupling $\omega_\gamma$
is  small. We may speak of a diluted random model.

The almost sure spectrum is denoted by
$\Sigma_{B,\lambda,\eta}$ and is contained in a union of intervals
$\I_n({B,\lambda})=[a_n(B,\lambda),b_n(B,\lambda)]\ni B_n$, $n
\in {\mathbb N}$. Moreover,  if
 ${\mathbb N} \ni N\le C
(B\lambda^2)^{-1}$ for some finite constant $C$, then
    \begin{equation}
\Sigma_{B,\lambda,\eta} \cap (-\infty, B_N+B] \subset
\bigcup_{n=1}^N \I_n({B,\lambda})\subset \bigcup_{n=1}^N
[B_n-C\lambda\sqrt{nB},
B_n+C\lambda\sqrt{nB}],\label{spectrum}
    \end{equation}
for some constant $C<\infty$. As a
consequence, for any integer $N\in {\mathbb N}$,  the first $N$
intervals $\I_n({B,\lambda})$, $n=1,\ldots,N$, are disjoint for
$\lambda$ small enough. More precisely,
  for any $B\in(0,\infty)$ there exists $\lambda_{\ast}$ such that for any
  $n\leq N$ and any $\lambda\in[0,\lambda_\ast)$ we have $\I_{n}({B,\lambda})\cap\I_{n+1}({B,\lambda})=\emptyset$,
  that is $b_{n}(B,\lambda) < a_{n+1}(B,\lambda)$. We denote by $\G_n(B,\lambda)
=(b_{n}(B,\lambda);a_{n+1}(B,\lambda))$ the $n$-th gap of the
spectrum. We say that the couple $(B,\lambda)$ respects the
\textit{ the disjoint band condition} if we have
\begin{equation}\label{DBC}
\G_n(B,\lambda)\neq\emptyset \text{ for any } n\leq N.
\end{equation}
It follows from \eqref{spectrum} that the disjoint band condition is satisfied if $\lambda\le C \sqrt{B/N}$ for some constant $C<\infty$.

\begin{thm}[\cite{DGR}]\label{thmresult}
Fix $N \in {\mathbb N}$. Let $H_{B,\lambda, \omega,\eta}$ be the
Hamiltonian described above, satisfying the disjoint band condition \eqref{DBC}. Then there exists a finite constant $\kappa_N>0$
(depending on $B$ and $N$) and $\Lambda=\Lambda(B,N)>0$, such that
for any $\lambda\in(0,\Lambda]$ and $\eta\in (0,c_{B,J}\lambda
|\log \lambda|^{-2}]$,  for all $n = 1,\cdots,N$, 
there exist dynamical mobility edges $E_{j,n}(B,\lambda)\in\mathcal{B}_n$, 
$j=1,2$, such that
\begin{align} 
\max_{j=1,2} \left \lvert E_{j,n}(B,\lambda)  - B_n  \right \rvert  
\le  \kappa_N\lambda^2.
\label{lambda20}
\end{align} 
(It is possible
that  ${E}_{1,n}(B,\lambda)=
{E}_{2,n}(B,\lambda)$, i.e., dynamical delocalization occurs at 
a single energy.)
\end{thm}

We further note that \cite{DGR} provides large classes of magnetic potentials for which the almost sure spectrum $\Sigma_{B,\lambda,\eta}$ is shown to contain  full intervals, of size $\mathcal{O}(\lambda)$ uniformly in $\eta>0$, centered at $B_n$, $n=1,\cdots,N$. We refer to \cite[Theorem 3.1]{DGR}.



\begin{thebibliography}{99}
%
%

\bibitem[AENSS]{AENSS}  Aizenman, M., Elgart, A., Naboko, S.,   
Schenker, J.,  Stolz, G., Moment analysis for localization in random Schr\"odinger 
operators. \textit{Inv. Math.}, \textbf{163} (2006), 343--413. 

\bibitem[AG]{AG} Aizenman, M.,   Graf, G.M.,
 Localization bounds for an electron gas. \textit{J. Phys. A: Math.
Gen.}, {\bf 31} (1998), 6783--6806.


\bibitem[An]{An}  Anderson, P.:  Absence of diffusion in certain random
lattices.  Phys. Rev. {\bf 109}, 1492-1505 (1958)


\bibitem[B]{Be}  Bellissard, J.:
 Ordinary quantum Hall effect and noncommutative cohomology.  Localization in disordered systems (Bad Schandau, 1986),  61-74, Teubner-Texte Phys., 16, Teubner, Leipzig, 1988


\bibitem[BES]{BES} Bellissard, J., van Elst, A., Schulz-Baldes, H., 
The non commutative geometry of the quantum Hall effect. \textit{J. Math. Phys.}, {\bf 35} (1994), 5373--5451.

\bibitem[DGR]{DGR} Dombrowski, N., Germinet, F., Raikov, G.,
Splitting of the Landau levels by magnetic perturbations and
Anderson transition in 2D-random magnetic media, preprint 2010, submitted.


\bibitem[GK1]{GK1} Germinet, F. and Klein, A.,
Bootstrap multiscale analysis and localization in random media, 
\textit{Comm. Math. Phys.},
\textbf{222} (1998), 415--448.

\bibitem[GK3]{GK3} Germinet, F. and Klein, A.,
 A characterization of the Anderson
metal-insulator transport transition,
\textit{Duke Mathematical Journal},
\textbf{124} (2004), 309--350.

\bibitem[GK5]{GKjsp} Germinet, F. and  Klein, A., New characterizations of
the region of complete localization for random Schr\"odinger operators.
\textit{J. Stat. Phys.}, {\bf 122} (2006), 73--94 

\bibitem[GKM]{GKM} Germinet, F,  Klein, A., Mandy, B.,
Dynamical delocalization in random Landau Hamiltonians with
unbounded random couplings, In: \textit{Spectral and Scattering
Theory for Quantum Magnetic Systems}, \textit{Contemp. Math.}, Amer. Math. Soc., Providence, RI, \textbf{500}
(2009),  87--100.


\bibitem[GKS1]{GKS} Germinet, F.,  Klein, A., Schenker, J.,
Dynamical delocalization in random Landau Hamiltonians,  \textit{ Annals
of Math.}, {\bf 166} (2007), 215--244.


\bibitem[GKS2]{GKS2} Germinet, F.,  Klein, A., Schenker, J.:
Quantization of the Hall conductance and delocalization in ergodic
Landau Hamiltonians, \textit{Rev. Math. Phys.}, {\bf 21} (2009), 1045--1080.

\bibitem[GhHK]{GHK} Ghribi, F., Hislop, P.D., Klopp, F.: Localization for Schr\"odinger
operators with random vector potentials, In: Contemporary
Mathematics, {\bf 447} (2007)  {\em Adventures in mathematical
physics}, Eds Germinet, Hislop, 123--138.


\bibitem[Kli]{vKli} von Klitzing, K, Dorda, G, Pepper, N.: New method for high-
accuracy determination of the fine structure constant based on quantized Hall 
resistance. Phys. Rev. Lett~{\bf 45}, 494 (1980).


\bibitem[RM]{RM} Rojas-Molina, C.,
 Characterization of the Anderson
metal-insulator transport transition for non ergodic operators and application,
in preparation.


%
%
%
%

\end{thebibliography}
\end{document}